\documentclass[english]{article}
\usepackage[T1]{fontenc}
\usepackage[latin9]{inputenc}
\usepackage{geometry}
\usepackage{xcolor}
\usepackage{pdfcolmk}
\usepackage{babel}
\usepackage{array}
\usepackage{float}
\usepackage{booktabs}
\usepackage{textcomp}
\usepackage{multirow}
\usepackage{amsmath}
\usepackage{graphicx}
\usepackage{setspace}
\usepackage[authoryear]{natbib}
\PassOptionsToPackage{normalem}{ulem}
\usepackage{ulem}
\usepackage[unicode=true,pdfusetitle,
 bookmarks=true,bookmarksnumbered=false,bookmarksopen=false,
 breaklinks=false,pdfborder={0 0 1},backref=false,colorlinks=true]
 {hyperref}
\usepackage[dot]{bibtopic}
\makeatletter
\providecommand{\tabularnewline}{\\}
\providecolor{lyxadded}{rgb}{0,0,1}
\providecolor{lyxdeleted}{rgb}{1,0,0}
\DeclareRobustCommand{\lyxsout}[1]{\ifx\\#1\else\sout{#1}\fi}
\usepackage[labelfont=bf]{caption}
\hypersetup{citecolor=blue}
\usepackage{graphicx} 
\usepackage{afterpage}
\usepackage{diagbox} % table diagonal box
\usepackage{chapterbib}
\usepackage[affil-it]{authblk}
\makeatother

\begin{document}

\title{Constructing energy accounts for WIOD 2016 release }

\author{Viktoras Kulionis\thanks{Email address: \texttt{viktoras.kulionis@ekh.lu.se}}} 
\affil{Department of Economic History,\\ School of Economics and Management,\\ Lund University, Sweden} 
\maketitle
\begin{abstract}
Most of today\textquoteright s products and services are made in global
supply chains. As a result a consumption of goods and services in
one country is associated with various environmental pressures all
over the world due to international trade. Advances in global multi-region
input-output models have allowed researchers to draw detailed, international
supply-chain connections between production and consumptions activities
and associated environmental impacts. Due to a limited data availability
there is little evidence about the more recent trends in global energy
footprint. In order to expand the analytical potential of the existing
WIOD 2016 dataset to wider range of research themes, this paper develops
energy accounts and presents the global energy footprint trends for
the period 2000-2014. 
\end{abstract}

\section{Introduction}

Addressing the problem of climate change has moved high up on the
governments\textquoteright{} agendas across the world. Effective strategies
to reduce country-specific impacts require accurate and reliable environmental
statistics. Such statistics should not only account for environmental
pressures occurring within the borders of a country but should also
allow to consider environmental pressures embodied in imports and
exports. 

This issue is of particular importance given that most of today\textquoteright s
products and services are no longer produced within a single country
and are made in global supply chains. This means that countries import
intermediate goods and raw materials, to which they add one or more
layers of value and sell the product either for final consumption
or for another producer who adds the next layer \citep{Tukker2013}.
Evidence suggest that the average number of border crossings in value
chain required for a product of one country to reach the final user
in another country is approximately 1.7 \citep{Muradov2016}. 

Normally environmental impacts are calculated following production-based
accounting (PBA) method. This method assigns the responsibility of
a specific factor (e.g. energy or CO$_2$) to a country where the
impact occurs. Following the rise in international trade and increasing
production fragmentation many scholars begun to discuss appropriate
ways to measure the responsibility for emissions and question the
effects of trade on the environment \citep{Tukker2013,Wiedmann_Lenzen2018}. 

One way to account for the factor content embodied in trade is to
use the consumption-based accounting (CBA). Significant attention
has been devoted to the use of consumption-based accounting principles
(also referred to as footprint) in the past few decades. Multi-regional
input-output (MRIO) analysis has proved to be an ideal tool for this
task. Recently, the availability of global multi-regional input-output
databases enabled researchers to draw detailed, global supply-chain
connections between production and consumption of goods and services. 

Multiple studies have shown that in the developed countries CO$_2$
and energy content embodied in imports is higher than in exports.
In contrast, for the developing countries the opposite is true, i.e.,
CO$_2$ and energy embodied in exports is higher than in imports.
Between 1995-2011, the share of total global environmental impacts
embodied in trade increased from 20\% to 29\% for energy use and from
19\% to 24\% for GHG emissions \citep{Wood2018}. 

While the MRIO models are a powerful tool for analysing the carbon
footprints of countries, their data and computational requirements
are often cited as barriers to timely, detailed and robust studies
\citep{Andrew2009}. Recent reviews of the main global MRIO initiatives
indicate that there are seven global MRIO databases \citep{Owen2015}.
Four of these (Eora, WIOD2013, EXIOBASE, GTAP) databases come with
the environmental extensions that permit environmental analyses (e.g.,
estimation of carbon or energy footprints). However, in some cases,
for instance, the WIOD database released in 2016 does not contain
environmental extensions. 

Monitoring and understanding global impacts associated with trade
of goods and services is essential for effective policy measures.
Global Databases with environmental extensions are necessary for this
task. This study aims to: i) demonstrate how data from the International
Energy Agency (IEA) can be used to construct energy accounts that
match the WIOD 2016 sectoral classification, ii) present detailed
comparison of WIOD2016 and WIOD2013 energy accounts, and iii) analyse
global energy production based accounts (PBA) and consumption based
accounts (CBA) for the period 2000-2014. 

\section{Data}

\subsection{Energy data}

Data for this study comes from two sources: i) International Agency
(IEA) and ii) World Input-Output Database (WIOD). \citet{IEA2016}
is the main source of energy data. Latest IEA 2017 edition provides
World Energy balances for 178 countries and regional aggregates over
the period 1960-2015 (OECD countries and regions) and 1971-2015 (non-OECD
countries and regions). For each year and country, energy balances
cover 67 products and 85 flows. For example, a flow \textquotedbl{}
iron \& steel\textquotedbl{} contains data of how much and what energy
product (e.g., coal, oil) iron and steel industry used during a specific
year. A final data extract from the IEA has the following dimensions:

\[
Year\times country\times flow\times product=14\times44\times63\times85
\]

It covers a 14 year period from 2000 to 20014, and contains data for
44 countries, 63 energy products and 85 flows.

\begin{table}[H]
\caption{\label{tab:Chapter4_IEA_EB}IEA energy balances, exemplified with
data for Germany 2014, (Mtoe)}
\centering{}\resizebox{\textwidth}{!}{{\footnotesize{}}%
\begin{tabular}{lccccr}
\toprule 
\multirow{2}{*}{{\footnotesize{}{\diagbox{\small{flow \textdownarrow}}{\small{product \textrightarrow}}}}} & \multicolumn{5}{c}{\textbf{\footnotesize{}Energy Product}}\tabularnewline
\cmidrule{2-6} 
 & {\footnotesize{}product $1$} & {\footnotesize{}product $2$} & {\footnotesize{}...} & {\footnotesize{}product $67$} & {\footnotesize{}Total}\tabularnewline
\midrule
\multicolumn{1}{l}{\textbf{\footnotesize{}TPES}} &  &  &  &  & \textbf{\footnotesize{}306}\tabularnewline
\multicolumn{1}{l}{{\footnotesize{}\hspace{1cm}Production}} & {\footnotesize{}...} & {\footnotesize{}...} & {\footnotesize{}...} & {\footnotesize{}...} & {\footnotesize{}120}\tabularnewline
{\footnotesize{}\hspace{1cm}Imports} & {\footnotesize{}...} & {\footnotesize{}...} & {\footnotesize{}...} & {\footnotesize{}...} & {\footnotesize{}246}\tabularnewline
{\footnotesize{}\hspace{1cm}Exports} & {\footnotesize{}...} & {\footnotesize{}...} & {\footnotesize{}...} & {\footnotesize{}...} & {\footnotesize{}-49}\tabularnewline
{\footnotesize{}\hspace{1cm}International marine bunkers} & {\footnotesize{}...} & {\footnotesize{}...} & {\footnotesize{}...} & {\footnotesize{}...} & {\footnotesize{}-2}\tabularnewline
{\footnotesize{}\hspace{1cm}International aviation bunkers} & {\footnotesize{}...} & {\footnotesize{}...} & {\footnotesize{}...} & {\footnotesize{}...} & {\footnotesize{}-8}\tabularnewline
{\footnotesize{}\hspace{1cm}Stock changes} & {\footnotesize{}...} & {\footnotesize{}...} & {\footnotesize{}...} & {\footnotesize{}...} & {\footnotesize{}0.3}\tabularnewline
{\footnotesize{}\hspace{1cm}Transfers} &  &  &  &  & {\footnotesize{}0.7}\tabularnewline
{\footnotesize{}\hspace{1cm}Statistical differences} &  &  &  &  & {\footnotesize{}0.3}\tabularnewline
\multicolumn{1}{l}{\textbf{\footnotesize{}Transformation processes}} &  &  &  &  & \textbf{\footnotesize{}-74}\tabularnewline
{\footnotesize{}\hspace{1cm}...} & {\footnotesize{}...} & {\footnotesize{}...} & {\footnotesize{}...} & {\footnotesize{}...} & \tabularnewline
\multicolumn{1}{l}{\textbf{\footnotesize{}Energy industry own use}} &  &  &  &  & \textbf{\footnotesize{}-16}\tabularnewline
{\footnotesize{}\hspace{1cm}...} & {\footnotesize{}...} & {\footnotesize{}...} & {\footnotesize{}...} & {\footnotesize{}...} & {\footnotesize{}...}\tabularnewline
\multicolumn{1}{l}{\textbf{\footnotesize{}Total final consumption}} &  &  &  &  & \textbf{\footnotesize{}216}\tabularnewline
{\footnotesize{}\hspace{1cm}Industry} &  &  &  &  & {\footnotesize{}55}\tabularnewline
{\footnotesize{}\hspace{1cm}...} & {\footnotesize{}...} & {\footnotesize{}...} & {\footnotesize{}...} & {\footnotesize{}...} & {\footnotesize{}...}\tabularnewline
{\footnotesize{}\hspace{1cm}Transport} &  &  &  &  & {\footnotesize{}55}\tabularnewline
{\footnotesize{}\hspace{1cm}...} & {\footnotesize{}...} & {\footnotesize{}...} & {\footnotesize{}...} & {\footnotesize{}...} & {\footnotesize{}...}\tabularnewline
{\footnotesize{}\hspace{1cm}Other} &  &  &  &  & {\footnotesize{}84}\tabularnewline
{\footnotesize{}\hspace{1cm}...} & {\footnotesize{}...} & {\footnotesize{}...} & {\footnotesize{}...} & {\footnotesize{}...} & {\footnotesize{}...}\tabularnewline
{\footnotesize{}\hspace{1cm}Non-energy use} &  &  &  &  & {\footnotesize{}22}\tabularnewline
\bottomrule
\end{tabular}}
\end{table}

\subsection{MRIO data}

Multi-regional input-output (MRIO) tables come from World Input Output
Database (WIOD), which contains WIOD 2013 release (WIOD13 hereafter)
and WIOD 2016 release (WIOD16 hereafter). WIOD13 version is a system
of MRIO tables, socioeconomic and environmental accounts \citep{Genty2012,Timmer2015,Timmer2016}.
It covers 35 industries and 41 countries/regions, including 27 EU
and 13 other major advanced and emerging economies, plus Rest of the
World (ROW) region over the period 1995-2011 (environmental accounts
only for 1995-2009). 

A more recent WIOD2016 database provides data for 56 industries and
44 countries (28 EU, 15 other major countries and ROW region) for
the period from 2000 to 2014 (see table \ref{tab:CH4_AppendixA_WIODCountry}
and table \ref{tab:CH4Appendix_WIOD_Sector} in Appendix A). It also
provides socio-economic accounts, but it lacks environmental accounts. 

The two databases overlap over the period from 2000 to 2009. WIOD2016
estimates are compared to WIOD2013 version over this period to test
for the accuracy of the WIOD2016 estimates. The aim is to provide
estimates that closely resemble those in WIOD2013 so that the two
databases could be linked to study the changes in environmental indicators
over an extended period: 1995-2014. This is a novel contribution of
this paper and could serve the scientific community in many ways.

\section{Methodology}

This section outlines the allocation procedure of the 85 flows of
the IEA energy balances into the corresponding WIOD16 sectors and
final demand categories. The allocation procedure have been outlined
in previous studies by \citet{Genty2012,Wood2015,Wiebe2016,Owen2017}.
The procedure to obtain energy accounts starting from energy balances
involves a series of steps. Each step with examples is explained below. 

\subsection{IEA Allocation Procedure}

\subsubsection{Step 1}

The IEA energy balances show the supply and the use of energy products
by industries and final use categories as in table \ref{tab:Chapter4_IEA_EB}.
This data allows to construct two energy extension vectors: one showing
energy use by industry and another showing energy supply of different
energy products (e.g. coal) by the source sector (e.g., Mining). The
two vectors are equivalent in size (energy supply = energy use), but
the allocation to industry sectors is different. Among the existing
databases GTAP and WIOD provide energy use vectors, Eora provides
energy-supply vectors, and EXIOBASE is the only database to provide
both energy vectors \citep{Owen2017}. There is little information
on the difference between the two vectors and the choice of which
energy extension vector to use when largely depends on the question
at hand. \citet{Owen2017} show that both energy extensions produce
very similar estimates of the overall energy CBA for the UK. However,
at a more detailed level, the results address different issues. For
instance, the energy-supply vector reveals how dependent the UK is
on the domestic energy supply, an issue that is of utmost importance
for energy security policy. On the other hand, the energy use vector
allows for the attribution of actual energy use to industry sectors,
which enables a better understanding of sectoral efficiency gains. 

In order to be consistent with WIOD13 energy accounts, this study
focuses on the construction of energy use instead of energy supply.
The very first step in deriving energy use accounts from the IEA energy
balances is to separate the use and the supply of energy products. 

Energy use consists of the total final consumption (Industry + transport
+ Other + Non-energy use); the aviation and marine bunkers; the energy
sector own use (with a changed algebraic sign) and transformation
processes (with a changed algebraic sign).

\subsubsection{Step 2}

The next step is to establish a correspondence key linking energy
balance items and WIOD16 industries plus households. An example of
a binary correspondence matrix is displayed in table . Zero value
\textquotedbl{}0\textquotedbl{} means no link and \textquotedbl{}1\textquotedbl{}
represents a link between the IEA flow and WIOD sector(s). The columns
containing only one entry represent one-to-one allocation, for example,
column $fl2$ is allocated to WIOD16 sector $s56$. The IEA flows
that contain multiple entries of \textquotedbl{}1\textquotedbl{} represent
one-to-many allocation. For instance, the IEA flow $fl1$ is allocated
to two WIOD16 sectors $s1$ and $s2$ and flow $fl85$ is split among
all WIOD sectors + households.

\begin{table}[H]
\caption{\label{tab:Chapter_ConcordanceExample}An example of a binary concordance
matrix}
\centering{}%
\begin{tabular}{>{\raggedright}p{4cm}ccccc}
\toprule 
 &  & \multicolumn{4}{c}{IEA energy flow}\tabularnewline
\cmidrule{3-6} 
 &  & $fl1$ & $fl2$ & ... & $fl85$\tabularnewline
\midrule
\multirow{5}{4cm}{WIOD16 \\
(56 sectors + households)} & $s1$ & 1 & 0 & ... & 1\tabularnewline
 & $s2$ & 1 & 0 & ... & 1\tabularnewline
 & ... & ... & ... & ... & ...\tabularnewline
 & $s56$ & 0 & 1 & ... & 1\tabularnewline
 & $hh$ & 0 & 0 & ... & 1\tabularnewline
\bottomrule
\end{tabular}
\end{table}

\subsubsection{Step 3}

While one-to-one allocation is a straightforward task one-to-many
allocation requires disaggregation of a specific IEA flow among several
WIOD16 sectors. The splitting key is the total input in monetary terms
from two WIOD16 energy related sectors: \textquotedblleft coke and
refined petroleum products\textquotedblright{} ($s10$) and \textquotedblleft Electricity,
gas, steam and air conditioning supply\textquotedblright{} ($s24$).
For instance, the splitting key to allocate IEA flow $f1$ among two
WIOD16 sectors $s1$ and $s2$ is $=[\begin{array}{cc}
12/16 & 4/16\end{array}]=[\begin{array}{cc}
0.75 & 0.25\end{array}]$ . This means that 75\% of IEA energy flow $f1$ is allocated to $s1$
and 25\% to $s2$.

\begin{table}[H]
\caption{\label{tab:CH5_spiltinvector}An example of a splitting key vector
(arbitrary numbers)}
\centering{}\resizebox*{\textwidth}{!}{%
\begin{tabular}{lcccccc}
\toprule 
 &  & \multicolumn{5}{c}{WIOD16 (\$)}\tabularnewline
\cmidrule{3-7} 
 &  & s1 & s2 & ... & s56 & HH\tabularnewline
\midrule
Coke and refined petroleum products & s10 & 5 & 1 & ... & 1 & 3\tabularnewline
Electricity, gas, steam and air conditioning supply & s24 & 7 & 3 & ... & 0 & 11\tabularnewline
\midrule
\textbf{Total} & \textbf{s10+s24} & \textbf{12} & \textbf{4} & \textbf{...} & \textbf{1} & \textbf{14}\tabularnewline
\bottomrule
\end{tabular}}
\end{table}

In a formal way the procedure in step 2 and step 3 can be written
as : 

\begin{equation}
\mathbf{M=\hat{a}C\widehat{aC}^{-1}}
\end{equation}

where $\mathbf{C}$ is a binary concordance matrix, $\mathbf{a}$
is a splitting vector and $\mathbf{M}$ is a mapping matrix between
IEA flow and WIOD16 sectors plus households. Using the information
from table \ref{tab:Chapter_ConcordanceExample} and table \ref{tab:CH5_spiltinvector}
this can be expressed in more detail as: 
\[
\underset{}{\mathbf{C}}=\left[\begin{array}{ccc}
1 & 0 & 1\\
1 & 0 & 1\\
0 & 1 & 1\\
0 & 0 & 1
\end{array}\right]\text{ }\underset{}{\mathbf{a}}=\left[\begin{array}{cccc}
12 & 4 & 1 & 14\end{array}\right]
\]
this yields:

\[
\underset{}{\mathbf{M}}=\left[\begin{array}{ccc}
0.75 & 0 & 0.39\\
0.25 & 0 & 0.13\\
0 & 1 & 0.03\\
0 & 0 & 0.45
\end{array}\right]
\]

\subsubsection{Step 4}

The above steps are combined to obtain the use of energy products
by WIOD16 sectors and final demand category using the following equation:

\begin{equation}
\mathbf{W16E=E\times M^{*}}
\end{equation}

Where $\mathbf{E}$ is the IEA energy use table as explained in Step
1 with dimension 63 x 5355 (63 products x 85 flows). This matrix is
obtained by diagonalising the 63x1 vector corresponding to each IEA
energy flow and stacking them horizontally. 

$\mathbf{M}^{*}$ is a 5355 x 57 energy use allocation matrix it is
obtained by modifying $\mathbf{M}$. Every column from $\mathbf{M}$
is transposed and replicated 63 times to match the energy product
dimension. $\mathbf{M}^{*}$ shows how much of each energy product
(corresponding to each energy flow) is used by each WIOD16 sector
plus households. 

$\mathbf{W16E}$ is the resulting energy use matrix with a dimension
63x57 representing the use of 63 energy products by 56 WIOD16 industries
plus households. The energy product dimension (63) has been further
aggregated to match WIOD13 classification of 27 energy products (See
Appendix for energy product detail). The final energy matrix is 27x57.
The above steps were repeated for all WIOD16 countries except Rest
of the World (RoW). For RoW energy use was estimated by taking IEA
World energy use and subtracting all energy use by WIOD16 countries.

\begin{table}[H]
\caption{\label{tab:Simplified-IEA-energy}Simplified IEA energy balance table,
(arbitrary numbers) }
\centering{}%
\begin{tabular}{lccc}
\toprule 
 &  & \multicolumn{2}{c}{IEA product}\tabularnewline
\cmidrule{3-4} 
 &  & $pr1$ & $pr2$\tabularnewline
\midrule
\multirow{3}{*}{IEA flow} & $fl1$ & 100 & 20\tabularnewline
 & $fl2$ & 4 & 2\tabularnewline
 & $fl2$ & 15 & 1\tabularnewline
\bottomrule
\end{tabular}
\end{table}

The procedure presented in step 4 can be illustrated using data from
step 2 and step 3. One additional piece of information needed for
the example is energy balance data i.e. $\mathbf{E}$. An example
of energy balance data is given in table \ref{tab:Simplified-IEA-energy}.
It displays the use of a specific energy product (e.g. oil, coal)
by a specific flow (e.g. transport). This information is presented
in a matrix form as :

\[
\underset{}{\mathbf{E}}=\left[\begin{array}{cccccc}
pr1 & 0 & pr1 & 0 & pr1 & 0\\
0 & pr2 & 0 & pr2 & 0 & pr2
\end{array}\right]=\left[\begin{array}{cccccc}
100 & 0 & 4 & 0 & 15 & 0\\
0 & 20 & 0 & 2 & 0 & 1
\end{array}\right]
\]

based on sample data from step 3 $\mathbf{M}^{*}$ is:

\[
\underset{}{\mathbf{M^{*}}}=\left[\begin{array}{cccc}
0.75 & 0.25 & 0 & 0\\
0.75 & 0.25 & 0 & 0\\
0 & 0 & 1 & 0\\
0 & 0 & 1 & 0\\
0.39 & 0.13 & 0.03 & 0.45\\
0.39 & 0.13 & 0.03 & 0.45
\end{array}\right]
\]

This matrix shows for $fl1$, 75\% of $pr1$ is allocated to $s1$
and 25\% to $s2$ and $pr2$ is allocated in the same way. For $fl2$
both $pr1$ and $pr2$ are allocated to $s3$.

Multiplying $\mathbf{E}$ and $\mathbf{M^{*}}$ yields:

\[
\underset{}{\mathbf{W16E}}=\left[\begin{array}{cccc}
80 & 27 & 4 & 6\\
15 & 5 & 2 & 0.5
\end{array}\right]
\]

elements in a first row display the use of $pr1$ by the four sectors,
the second row show the use of $pr2$. For instance, $s1$ uses 80
units of $pr1$ and 15 units of $pr2$.

\subsubsection{Step 5: Accuracy}

The accuracy of WIOD16 energy use estimates was evaluated by measuring
the difference between WIOD13 energy and WIOD16 energy. \citet{Steen-Olsen2014}
have used a similar approach to estimate MRIO aggregation error. The
relative error $\varepsilon$ between WIOD13 $(\mathbf{W13})$ and
WIOD16 $(\mathbf{W16})$ for a given year $t$ and country $r$ is
defined as:

\[
\varepsilon_{t}^{r}=\frac{\mathbf{W16}_{t}^{r}-\mathbf{W13}_{t}^{r}}{\mathbf{W13}_{t}^{r}}
\]

where $\mathbf{W13}$ and $\mathbf{W16}$ is total energy use (from
a production perspective) for WIOD13 and WIOD16 respectively. 

\subsubsection{Step 6: Calibration}

The final step is to calibrate WIOD16 estimates so that they match
those of WIOD13 for the year where the two databases overlap, i.e.,
2000-2009. It is important to note that while sectoral detail does
not match between the two databases energy product detail is the same,
i.e. in WIOD13 energy use for a single country is given by $35\times27$
energy matrix and WIOD16 $56\times27$, hence the total energy use
by energy product is given by $1\times27$ vector. The calibration
was performed in two steps. First, total energy use by product in
WIOD16 ($\mathbf{W16E}$) is divided by total product use in WIOD13
($\mathbf{W13E}$) as:

\[
\alpha=(\mathbf{W16E}_{t}^{r}\mathbf{i)'}\widehat{\mathbf{W13E}_{t}^{r}\mathbf{i}}^{-1}
\]

Where $\mathbf{W16E}$ is $56\times27$ WIOD16 energy use matrix,
$\mathbf{W13E}$ is $35\times27$ energy use by product, $\mathbf{i}$
is a vector of ones used for summation, $\alpha$is $1\times27$ vector
that shows over/under estimation of a particular energy product. The
second step involves adjusting WIOD16 energy accounts as:

\[
\mathbf{W16E}^{calibrated}=\widehat{\alpha}\mathbf{W16E}
\]

Here it is assumed that under/over estimation of a particular energy
product is equally distributed among all sectors. For instance, if
coal use in WIOD16 is found to be underestimated by 2\% then for every
industry that uses coal its consumption is raised by 2\%. The calibration
strategy is applied for the years where the two datasets overlap,
i.e., 2000-2009. 

For the period 2010-2014 an additional step was required to calibrate
the estimates. It involved extrapolation of under/over estimation
data from previous years using a 5-year moving average. 

In order to show the scale of adjustments between the two databases,
the results are provided for energy use before and after the calibration.
While the energy CBA calculations are performed only using the calibrated
data.

\subsection{Calculation of Energy Footprint}

A standard environmentally extended Leontief model is applied to calculate
energy footprints for WIOD13 and WIOD16. The basic Leontief model
can be expressed as: 

\[
\mathbf{x}=(\mathbf{I-A})^{-1}\mathbf{Y}=\mathbf{LY}
\]

where $\mathbf{x}$ is the vector of output, $\mathbf{A}$ is the
matrix of technical coefficients, $\mathbf{Y}$ is the matrix of final
demands and $\mathbf{(I-A)^{-1}=L}$ is the total requirement matrix
representing interdependencies between industries. The IO model in
equation 1 is extended to incorporate energy use as:

\[
\mathbf{E}=\mathbf{e'L}\mathbf{Y}
\]

where $\mathbf{E}$ is the total energy requirements from consumption
perspective (CBA) and $\mathbf{e}$ is the direct energy intensity
vector representing energy use per unit of output for a given country.s

\section{Results }

\subsection{WIOD16 allocation results }

The difference between WIOD16 energy use estimates in comparison with
WIOD13 for selected years and the average for the period 2000-2009
are presented in table \ref{tab:Chapter4_Tbl1}. The results indicate
that for most countries WIOD16 and WIOD13 results vary between 1 and
4 per cent and in most cases the difference is positive. For the world
total, the results are higher on average by 4.1\% implying that WIOD16
energy use estimates are on average higher than WIOD13. However, there
are also some exceptions, e.g., Denmark, the Netherlands and Germany. 

For Denmark, Malta, Belgium and Luxembourg the estimates display greater
discrepancies and vary between 10-20\%. For Denmark and Luxembourg
the results are underestimated and for Malta and Belgium overestimated.
For China and Austria WIOD16, energy use estimates are on average
7-9 \% larger than WIOD13. For these countries, the results are less
accurate (assuming WIOD13 is a correct measure) than for the rest
of the sample, but they are precise (i.e. over/underestimation is
similar over the years).

Switzerland, Croatia and Norway were not included in the WIOD13 release,
and therefore it was not possible to present the estimation error
for these countries.

\afterpage{ 
\thispagestyle{empty}
\begin{table}[H]
\centering{}\caption{\label{tab:Chapter4_Tbl1}Estimation error of WIOD 2016 Energy use
accounts}
\resizebox*{!}{\textheight}{%
\begin{tabular}{lrrrr}
\toprule 
 & {\small{}2000 } & {\small{}2005} & {\small{}2009} & {\small{}2000-2009}\tabularnewline
 & $\varepsilon_{00}(\%)$ & $\varepsilon_{05}(\%)$ & $\varepsilon_{09}(\%)$  & $\left|\varepsilon_{00-09}\right|(\%)$\tabularnewline
\midrule
{\small{}Denmark} & {\small{}-11.0} & {\small{}-19.9} & {\small{}-23.2} & {\small{}18.8}\tabularnewline
{\small{}Malta} & {\small{}14.4} & {\small{}8.2} & {\small{}29.3} & {\small{}14.7}\tabularnewline
{\small{}Belgium} & {\small{}9.1} & {\small{}13.1} & {\small{}10.7} & {\small{}13.8}\tabularnewline
{\small{}Luxembourg} & {\small{}-10.4} & {\small{}-9.2} & {\small{}-18.5} & {\small{}11.3}\tabularnewline
{\small{}China} & {\small{}7.3} & {\small{}10.1} & {\small{}8.6} & {\small{}9.1}\tabularnewline
{\small{}Austria} & {\small{}7.7} & {\small{}9.1} & {\small{}7.0} & {\small{}8.3}\tabularnewline
{\small{}Slovakia} & {\small{}4.9} & {\small{}5.9} & {\small{}5.6} & {\small{}5.4}\tabularnewline
{\small{}Rest of World} & {\small{}3.4} & {\small{}4.4} & {\small{}5.6} & {\small{}4.5}\tabularnewline
{\small{}Spain} & {\small{}5.0} & {\small{}4.2} & {\small{}4.4} & {\small{}4.4}\tabularnewline
{\small{}Finland} & {\small{}5.5} & {\small{}5.0} & {\small{}3.3} & {\small{}4.4}\tabularnewline
{\small{}Netherlands} & {\small{}-6.4} & {\small{}-4.3} & {\small{}-1.0} & {\small{}3.9}\tabularnewline
{\small{}Taiwan} & {\small{}-1.9} & {\small{}-3.2} & {\small{}-5.4} & {\small{}3.3}\tabularnewline
{\small{}Brazil} & {\small{}3.2} & {\small{}3.0} & {\small{}2.7} & {\small{}3.2}\tabularnewline
{\small{}Ireland} & {\small{}1.2} & {\small{}-2.8} & {\small{}-8.8} & {\small{}3.2}\tabularnewline
{\small{}Romania} & {\small{}1.8} & {\small{}4.4} & {\small{}2.1} & {\small{}3.1}\tabularnewline
{\small{}Czech Republic} & {\small{}3.0} & {\small{}2.9} & {\small{}2.5} & {\small{}3.0}\tabularnewline
{\small{}Greece} & {\small{}5.5} & {\small{}-2.9} & {\small{}-2.3} & {\small{}2.8}\tabularnewline
{\small{}Bulgaria} & {\small{}2.9} & {\small{}2.8} & {\small{}1.3} & {\small{}2.6}\tabularnewline
{\small{}Latvia} & {\small{}-1.4} & {\small{}-1.7} & {\small{}5.3} & {\small{}2.2}\tabularnewline
{\small{}Cyprus} & {\small{}1.2} & {\small{}3.1} & {\small{}-1.4} & {\small{}1.9}\tabularnewline
{\small{}Russia} & {\small{}2.1} & {\small{}1.7} & {\small{}1.1} & {\small{}1.7}\tabularnewline
{\small{}Poland} & {\small{}1.6} & {\small{}1.6} & {\small{}1.5} & {\small{}1.6}\tabularnewline
{\small{}Sweden} & {\small{}2.0} & {\small{}1.2} & {\small{}1.7} & {\small{}1.6}\tabularnewline
{\small{}Estonia} & {\small{}-2.6} & {\small{}-1.2} & {\small{}-0.1} & {\small{}1.4}\tabularnewline
{\small{}France} & {\small{}1.6} & {\small{}1.4} & {\small{}1.1} & {\small{}1.4}\tabularnewline
{\small{}Portugal} & {\small{}1.7} & {\small{}1.0} & {\small{}1.2} & {\small{}1.4}\tabularnewline
{\small{}Great Britain} & {\small{}1.0} & {\small{}1.1} & {\small{}1.2} & {\small{}1.3}\tabularnewline
{\small{}Italy} & {\small{}0.3} & {\small{}1.7} & {\small{}1.9} & {\small{}1.3}\tabularnewline
{\small{}Canada} & {\small{}0.6} & {\small{}1.5} & {\small{}1.6} & {\small{}1.2}\tabularnewline
{\small{}Hungary} & {\small{}0.7} & {\small{}1.4} & {\small{}0.8} & {\small{}1.2}\tabularnewline
{\small{}Australia} & {\small{}0.4} & {\small{}-1.6} & {\small{}1.8} & {\small{}1.1}\tabularnewline
{\small{}Germany} & {\small{}-0.2} & {\small{}-0.9} & {\small{}-1.0} & {\small{}1.0}\tabularnewline
{\small{}Mexico} & {\small{}-0.6} & {\small{}-0.1} & {\small{}0.5} & {\small{}1.0}\tabularnewline
{\small{}Indonesia} & {\small{}0.2} & {\small{}-0.4} & {\small{}0.5} & {\small{}0.9}\tabularnewline
{\small{}India} & {\small{}-0.9} & {\small{}-0.2} & {\small{}1.0} & {\small{}0.7}\tabularnewline
{\small{}South Korea} & {\small{}0.9} & {\small{}0.6} & {\small{}-0.4} & {\small{}0.7}\tabularnewline
{\small{}Lithuania} & {\small{}0.1} & {\small{}1.1} & {\small{}1.5} & {\small{}0.7}\tabularnewline
{\small{}Japan} & {\small{}-0.3} & {\small{}-0.5} & {\small{}-1.0} & {\small{}0.6}\tabularnewline
{\small{}Slovenia} & {\small{}0.2} & {\small{}0.1} & {\small{}1.7} & {\small{}0.4}\tabularnewline
{\small{}Turkey} & {\small{}-0.1} & {\small{}-0.1} & {\small{}-1.0} & {\small{}0.4}\tabularnewline
{\small{}United States } & {\small{}0.2} & {\small{}0.3} & {\small{}-0.5} & {\small{}0.2}\tabularnewline
{\small{}Switzerland} & {\small{}n/a} & {\small{}n/a} & {\small{}n/a} & {\small{}n/a}\tabularnewline
{\small{}Croatia} & {\small{}n/a} & {\small{}n/a} & {\small{}n/a} & {\small{}n/a}\tabularnewline
{\small{}Norway} & {\small{}n/a} & {\small{}n/a} & {\small{}n/a} & {\small{}n/a}\tabularnewline
{\small{}World Total} & {\small{}2.7} & {\small{}4.5} & {\small{}5.1} & {\small{}4.1}\tabularnewline
\bottomrule
\end{tabular}}
\end{table}
}

\quad{}

\subsection{PBA and CBA results WIOD13 vs WIOD16}

\begin{figure}[H]
\caption{\label{fig:CH4_fig1}PBA and CBA energy use for the USA, WIOD13 vs
WIOD16 and WIOD16C }

\includegraphics[width=1\textwidth]{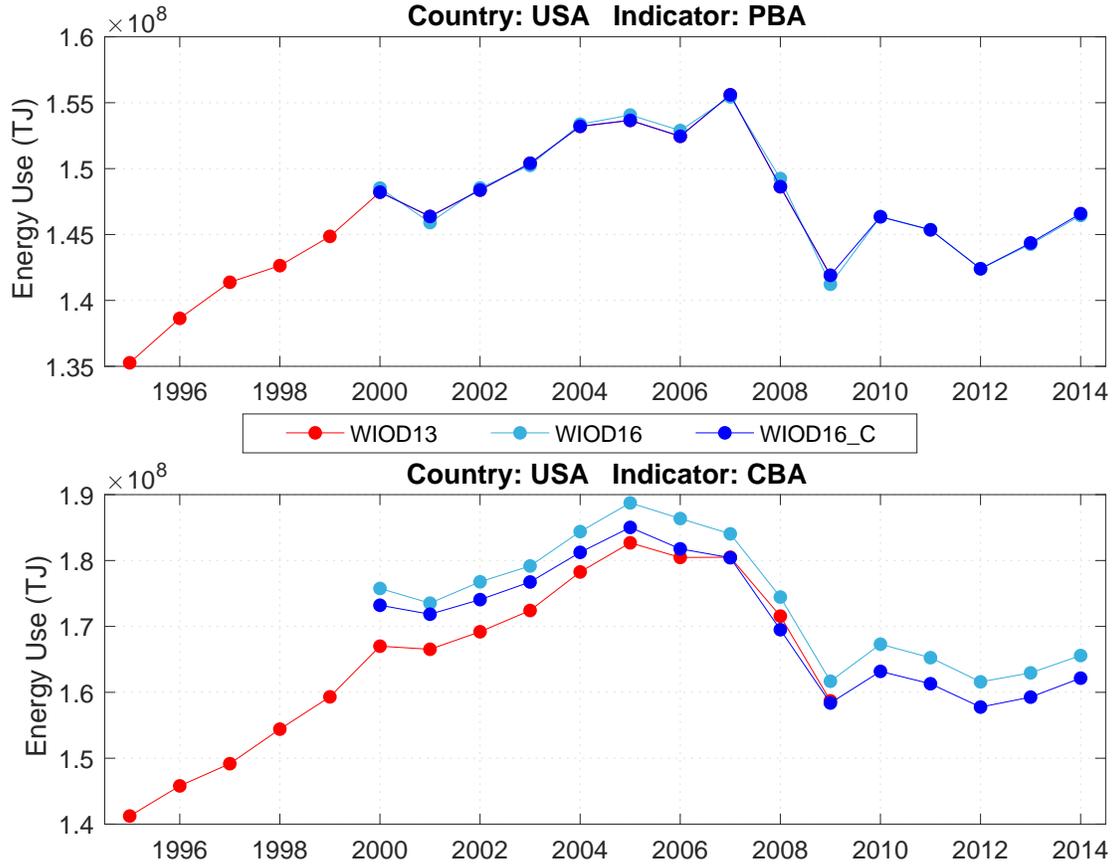}
\end{figure}

\begin{figure}[H]
\caption{\label{fig:CH4_fig2}PBA and CBA energy use for China, WIOD13 vs WIOD16
and WIOD16C }

\includegraphics[width=1\textwidth]{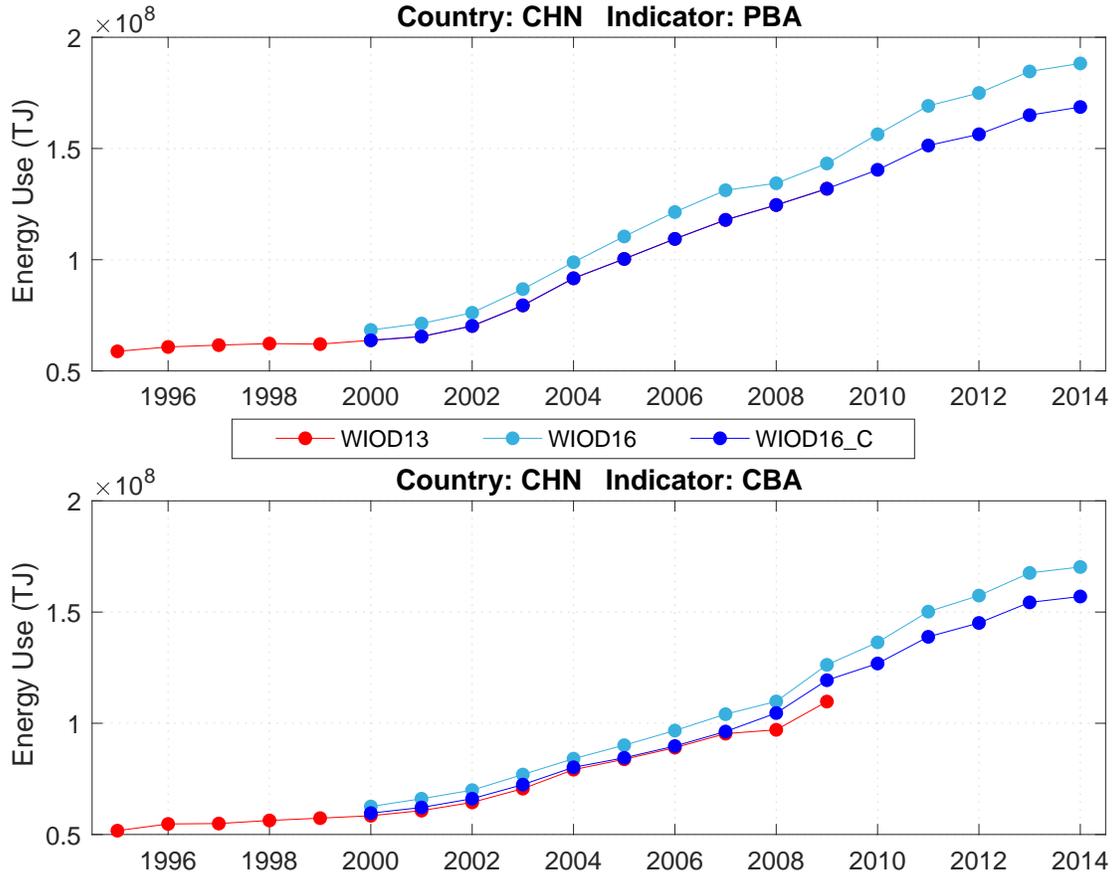}
\end{figure}

\begin{figure}[H]
\caption{\label{fig:CH4_fig3}PBA and CBA energy use for the Japan, WIOD13
vs WIOD16 and WIOD16C }

\includegraphics[width=1\textwidth]{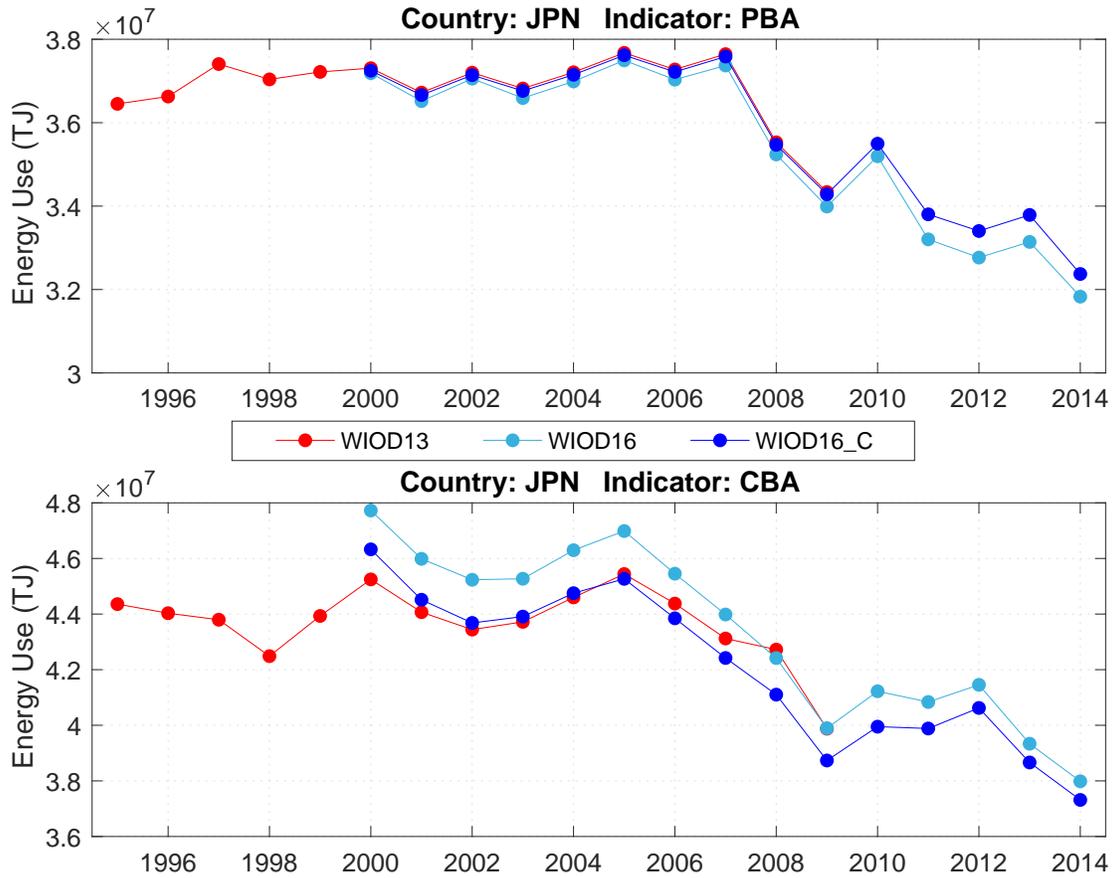}
\end{figure}

\begin{figure}[H]
\caption{\label{fig:CH4_fig4}PBA and CBA energy use for the Germany, WIOD13
vs WIOD16 and WIOD16C }

\includegraphics[width=1\textwidth]{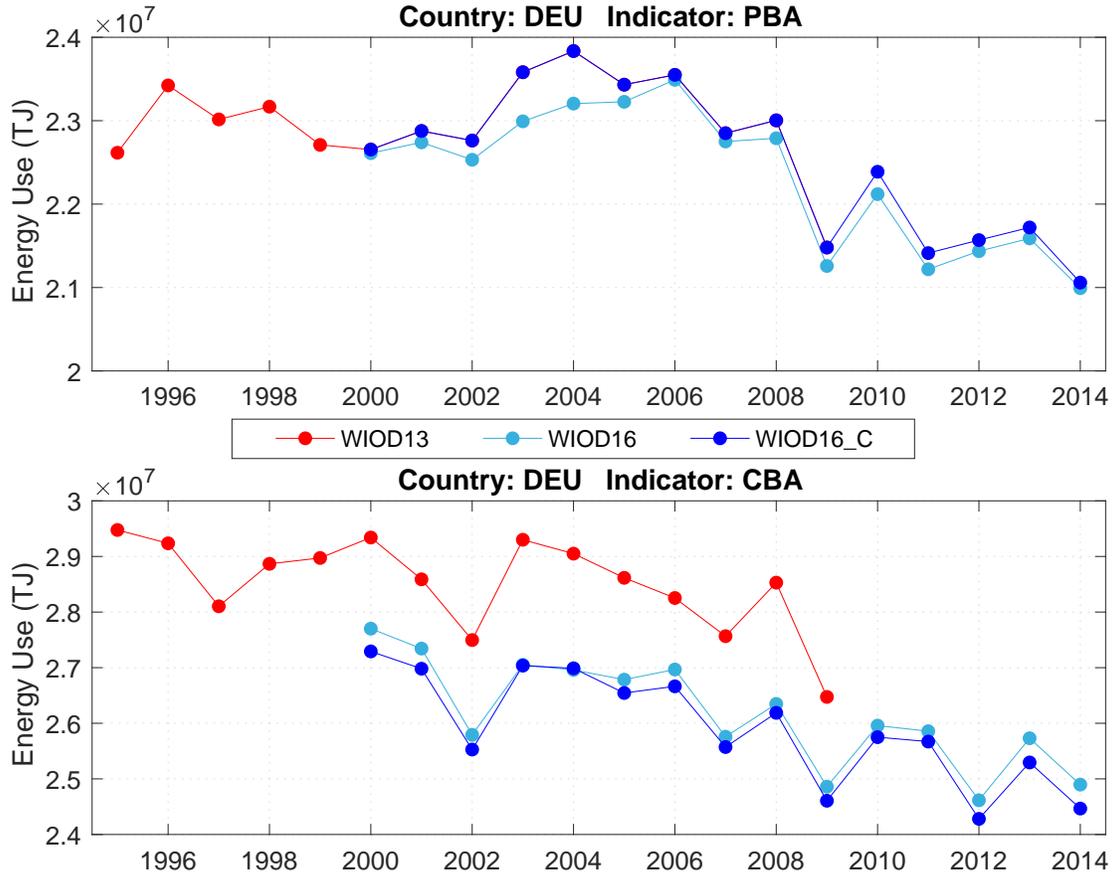}
\end{figure}

WIOD16 energy use estimates were calibrated to match WIOD13 estimates
(the procedure explained in step 6). The calibrated energy accounts
are labeled as WIOD16C. Data from all three (WIOD13, WIOD16, WIOD16C)
energy accounts have been used to calculate PBA and CBA for the period
2000 to 2014 (1995-2009 for WIOD13). To show yearly variations between
different estimates the results are displayed for four selected countries
(China, Germany, Japan and the US) in figures \ref{fig:CH4_fig1},\ref{fig:CH4_fig2},\ref{fig:CH4_fig3},\ref{fig:CH4_fig4}.
The two databases overlap from 2000 to 2009, so this period can be
used to study the differences between WIOD13 and WIOD16 and WIOD16\_C.
It is important to note that PBA indicator for WIOD\_C and WIOD13
is the same (or very close) due to calibration but CBA can differ,
for example, due to a greater sectoral and country detail. 

Figure \ref{fig:CH4_fig1} display CBA and PBA results for the USA.
PBA results are virtually the same when calculated using WIOD13 and
WIOD16. On the other hand, CBA results are larger when using WIOD16
especially during the period 2000-2006. Finally, we can see that energy
use has stabilised in the US after 2008 for both PBA and CBA measures.
The are no difference between WIOD16C and WIOD16 for the PBA indicator
and for CBA the results are when using WIOD16C, but between 2000-2006
they are still higher than WIOD13.

The same results are displayed for China in figure \ref{fig:CH4_fig2}.
Here, we can see that WIOD16 results are higher for both PBA and CBA
measures, but they follow the same trend as WIOD13. The results for
the period after 2009 show that energy use in China continues to increase.
WIOD16C results show that with calibrated data CBA measure is almost
identical to WIOD13. 

The results for Japan are displayed in figure \ref{fig:CH4_fig3}.
In general, the results for Japan are similar to those of the USA.
PBA energy use is virtually the same when calculated using WIOD13
and WIOD16. Whereas, CBA is higher when calculated with WIOD16 than
with WIOD13. From 2009 PBA and CBA has declined in Japan. WIODC closely
follow WIOD13 for CBA indicator until 2005 after which WIODC gives
a lower CBA estimate than WIOD13. 

The results for Germany displayed in figure \ref{fig:CH4_fig4} show
a different story. PBA estimates are virtually the same according
to both WIOD13 and WIOD16 calculations. CBA results are different
in the sense that WIOD16 display lower values than WIOD13 which is
opposite to the deviations seen for the US and Japan. For CBA indicator
WIOD16C results are very similar to WIOD16 prior calibration 

\subsection{PBA and CBA results for WIOD16}

How did energy footprint develop after 2009? and What are the effects
of a greater sectoral (56 vs 35) and country detail (44 vs 41) for
CBA estimates? To address these questions WIOD16C energy use estimates
are presented in table 2 for all countries.

\begin{figure}[H]
\caption{\label{fig:CH4_Boxplot}Difference between WIOD13 and WIO16C energy
footprint (CBA), 2000-2009 }

\includegraphics[width=1\textwidth]{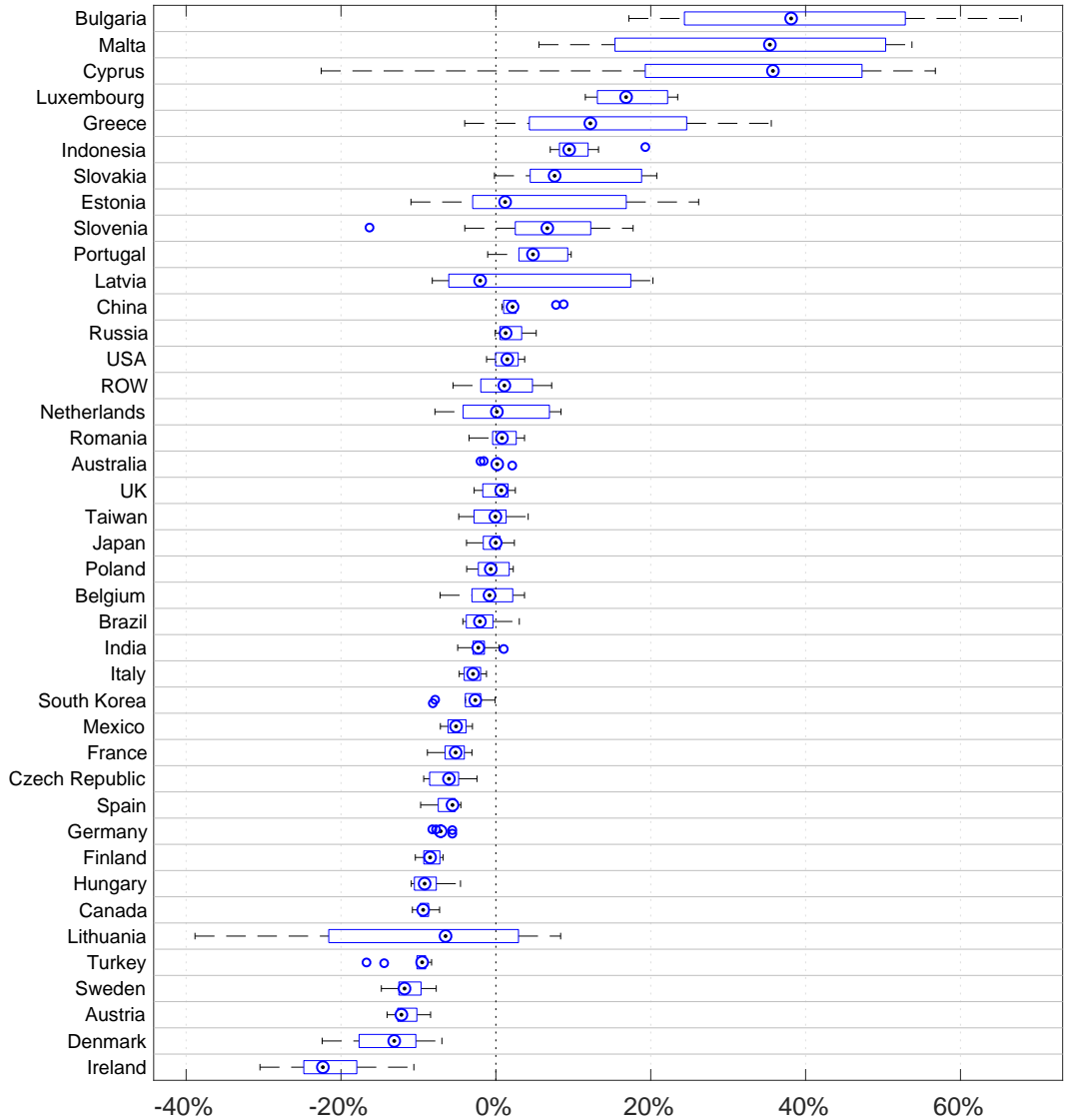}
\end{figure}

The difference between WIOD13 and WIOD16C energy footprints (CBA)
for the period 2000 - 2009 are shown in figure \ref{fig:CH4_Boxplot}.
For countries at the top of the figure CBA results are higher when
calculated using WIOD16C, and for countries, at the bottom of the
graph, the results are lower. 

Bulgaria, Malta and Cyprus stand out as outliers in this sample. WIOD16C
CBA estimates are on average 30-40\% higher than in WIOD13. These
countries also display a high degree of variation (0 \textendash{}
55\%) in the results. This implies that in some years the results
are quite similar while in others they differ substantially. High
degree of variation in the results is also visible for Greece, Slovakia,
Estonia, Slovenia, Latvia and Lithuania. 

Other countries that have higher CBA in WIOD16C the results fall in
the 0-10\% range. For instance, for China, Russia, and the US CBA
estimates are on average 1.5-3\% higher. 

Another set of countries including the Netherlands, Romania, Australia,
the UK, Taiwan, Japan and Poland do not show significant differences
between different databases. The results for these countries are within
\textpm 1\%. 

For the remaining countries at the bottom part of figure \ref{fig:CH4_Boxplot}
CBA results are lower when calculated using WIOD16C. For most countries,
the estimates vary between 0-10\%. A few notable exceptions are Sweden,
Austria, Denmark and Ireland. For these countries, WIOD16C CBA estimates
are more than 10\% lower compared to WIOD13. The majority of the countries
with lower CBA estimates are the EU countries.

The differences between the two databases can occur due to several
reasons. First, more detailed sectoral classification (from 35 to
56) can lead to lower estimates if disaggregated sectors (in WIOD16C)
have different energy intensities and imports occur predominantly
from a sector with a lower intensity. Second, a more detailed country
classification can lead to the same outcome if imports come from a
country with lower energy intensities than the rest of the world (ROW)
aggregate. Finally, the differences in how IO tables and Energy accounts
have been compiled also play a role.

\quad{}

\subsubsection{Global Energy footprint 2000-2014}

Estimates of CBA and PBA on per capita basis for the years 2000 vs
2014 and the per cent change over the period 2000-2014 are presented
in Table \ref{tab:Chapter4_Tbl2} for all countries covered by WIOD16.
The results for both PBA and CBA show substantial variations across
countries. 

\quad{}

\afterpage{ 
\thispagestyle{empty}
\begin{table}[H]
\centering{}\caption{\label{tab:Chapter4_Tbl2}PBA vs CBA, 2000 and 2014 (GJ/per capita)}
\resizebox*{!}{\textheight}{%
\begin{tabular}{lrrrrrrr}
\toprule 
 & \multicolumn{3}{c}{\textbf{PBA}} &  & \multicolumn{3}{c}{\textbf{CBA}}\tabularnewline
\cmidrule{2-4} \cmidrule{6-8} 
 & 2000 & 2014 & $\bigtriangleup\%$  &  & 2000 & 2014 & $\bigtriangleup\%$ \tabularnewline
\midrule
{\small{}Australia} & {\small{}333} & {\small{}335} & {\small{}0.4} &  & 342 & 455 & 33,0\tabularnewline
{\small{}Austria} & {\small{}224} & {\small{}230} & {\small{}2.7} &  & 302 & 296 & -1,9\tabularnewline
{\small{}Belgium} & {\small{}429} & {\small{}354} & {\small{}-17.5} &  & 413 & 345 & -16,6\tabularnewline
{\small{}Bulgaria} & {\small{}151} & {\small{}168} & {\small{}11.1} &  & 167 & 142 & -15,1\tabularnewline
{\small{}Brazil} & {\small{}75} & {\small{}97} & {\small{}29.6} &  & 82 & 107 & 30,5\tabularnewline
{\small{}Canada} & {\small{}556} & {\small{}527} & {\small{}-5.3} &  & 430 & 484 & 12,4\tabularnewline
{\small{}Switzerland} & {\small{}221} & {\small{}196} & {\small{}-11.2} &  & 372 & 337 & -9,3\tabularnewline
{\small{}China} & {\small{}51} & {\small{}123} & {\small{}143.6} &  & 47 & 115 & 143,0\tabularnewline
{\small{}Cyprus} & {\small{}178} & {\small{}102} & {\small{}-42.8} &  & 194 & 233 & 20,2\tabularnewline
{\small{}Czech Republic} & {\small{}241} & {\small{}246} & {\small{}1.8} &  & 218 & 212 & -2,6\tabularnewline
{\small{}Germany} & {\small{}276} & {\small{}260} & {\small{}-5.6} &  & 332 & 302 & -8,9\tabularnewline
{\small{}Denmark} & {\small{}309} & {\small{}267} & {\small{}-13.5} &  & 337 & 313 & -7,2\tabularnewline
{\small{}Spain} & {\small{}210} & {\small{}185} & {\small{}-11.7} &  & 233 & 188 & -19,3\tabularnewline
{\small{}Estonia} & {\small{}201} & {\small{}276} & {\small{}37.3} &  & 287 & 277 & -3,5\tabularnewline
{\small{}Finland} & {\small{}443} & {\small{}437} & {\small{}-1.4} &  & 396 & 381 & -3,8\tabularnewline
{\small{}France} & {\small{}279} & {\small{}225} & {\small{}-19.5} &  & 313 & 274 & -12,4\tabularnewline
{\small{}Great Britain} & {\small{}260} & {\small{}184} & {\small{}-29.1} &  & 324 & 269 & -17,0\tabularnewline
{\small{}Greece} & {\small{}216} & {\small{}217} & {\small{}0.5} &  & 349 & 223 & -36,0\tabularnewline
{\small{}Croatia} & {\small{}142} & {\small{}125} & {\small{}-11.7} &  & 147 & 135 & -8,6\tabularnewline
{\small{}Hungary} & {\small{}158} & {\small{}156} & {\small{}-1.3} &  & 162 & 145 & -10,4\tabularnewline
{\small{}Indonesia} & {\small{}43} & {\small{}48} & {\small{}13.3} &  & 40 & 61 & 51,8\tabularnewline
{\small{}India} & {\small{}25} & {\small{}40} & {\small{}61.0} &  & 24 & 36 & 52.4\tabularnewline
{\small{}Ireland} & {\small{}221} & {\small{}170} & {\small{}-22.7} &  & 272 & 246 & -9.4\tabularnewline
{\small{}Italy} & {\small{}221} & {\small{}167} & {\small{}-24.6} &  & 284 & 198 & -30.4\tabularnewline
{\small{}Japan} & {\small{}294} & {\small{}257} & {\small{}-12.6} &  & 365 & 295 & -19.2\tabularnewline
{\small{}South Korea} & {\small{}328} & {\small{}407} & {\small{}24.0} &  & 278 & 318 & 14.6\tabularnewline
{\small{}Lithuania} & {\small{}173} & {\small{}229} & {\small{}32.2} &  & 236 & 218 & -7.9\tabularnewline
{\small{}Luxembourg} & {\small{}398} & {\small{}367} & {\small{}-7.6} &  & 675 & 639 & -5.3\tabularnewline
{\small{}Latvia} & {\small{}90} & {\small{}115} & {\small{}27.6} &  & 170 & 184 & 8.2\tabularnewline
{\small{}Mexico} & {\small{}101} & {\small{}97} & {\small{}-4.5} &  & 116 & 109 & -5.7\tabularnewline
{\small{}Malta} & {\small{}150} & {\small{}181} & {\small{}20.2} &  & 407 & 236 & -42.1\tabularnewline
{\small{}Netherlands} & {\small{}478} & {\small{}440} & {\small{}-8.0} &  & 332 & 281 & -15.4\tabularnewline
{\small{}Norway} & {\small{}508} & {\small{}449} & {\small{}-11.6} &  & 426 & 470 & 10.3\tabularnewline
{\small{}Poland} & {\small{}149} & {\small{}161} & {\small{}8.2} &  & 156 & 159 & 2.0\tabularnewline
{\small{}Portugal} & {\small{}171} & {\small{}160} & {\small{}-6.4} &  & 231 & 174 & -24.4\tabularnewline
{\small{}Romania} & {\small{}107} & {\small{}107} & {\small{}0.6} &  & 94 & 117 & 24.7\tabularnewline
{\small{}Russia} & {\small{}297} & {\small{}366} & {\small{}23.2} &  & 176 & 272 & 54.3\tabularnewline
{\small{}Slovakia} & {\small{}216} & {\small{}196} & {\small{}-9.2} &  & 215 & 202 & -6.2\tabularnewline
{\small{}Slovenia} & {\small{}169} & {\small{}164} & {\small{}-2.8} &  & 295 & 236 & -20.2\tabularnewline
{\small{}Sweden} & {\small{}420} & {\small{}368} & {\small{}-12.4} &  & 391 & 356 & -9.0\tabularnewline
{\small{}Turkey} & {\small{}76} & {\small{}95} & {\small{}25.1} &  & 95 & 116 & 22.5\tabularnewline
{\small{}Taiwan} & {\small{}289} & {\small{}343} & {\small{}18.6} &  & 287 & 238 & -17.1\tabularnewline
{\small{}United States } & {\small{}525} & {\small{}461} & {\small{}-12.2} &  & 614 & 510 & -16.8\tabularnewline
{\small{}Rest of World} & {\small{}65} & {\small{}71} & {\small{}8.6} &  & 53 & 67 & 28.2\tabularnewline
\bottomrule
\end{tabular}}
\end{table}
}

Developed economies, in general, have higher PBA and CBA than developing
countries. The USA, Canada and many European countries have PBA and
CBA of more than 350 GJ/per capita. In contrast, India has about ten
times lower PBA and CBA accounting for roughly 35 GJ/per capita. China
had the highest PBA and CBA growth in the sample, both measures increased
by 143\% between 2000 and 2014, but the levels in 2014 are still less
than half of the values for developed countries. 

\begin{figure}[H]
\caption{\label{fig:CH4_PBA_CBA_global}PBA and CBA, 2000-2014 selected countries/regions }

\includegraphics[width=1\textwidth]{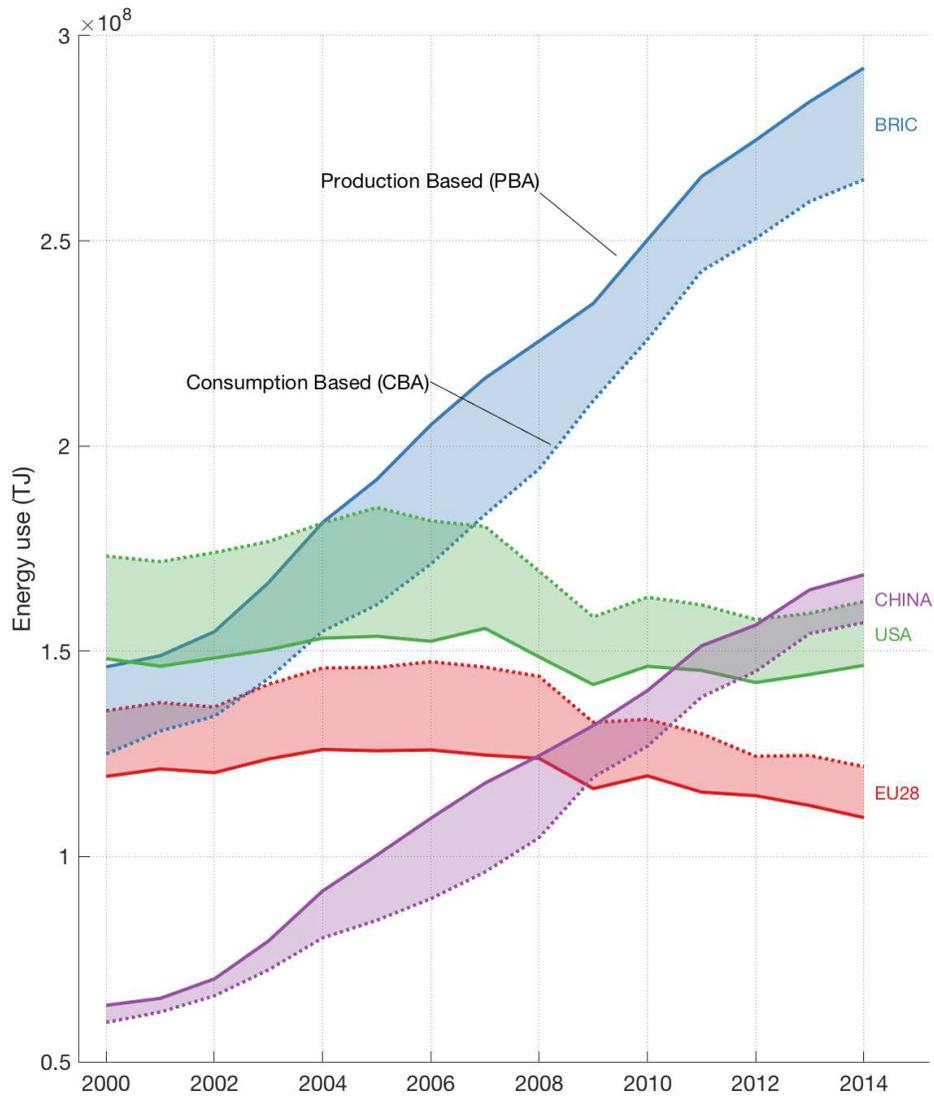}
\end{figure}

Figure \ref{fig:CH4_PBA_CBA_global} display PBA (solid line) and
CBA (dashed line) results over the period 2000-2014 for selected countries
and regions. The area between the two lines represents net import
(net export) of energy embodied in trade (aka BEET). The solid line
above the dashed line implies that a country/region is a net importer
of energy and the dashed line above the solid implies that a country
is a net exporter of energy.

BRIC and China are net exporters of energy. More energy is embodied
in exports of goods and services than in imports. For the USA and
the EU28, the result is the opposite. Furthermore, BRIC and China
display an increasing PBA and CBA trend, while the USA and the EU28
show stable or declining trend (especially for the EU28). 

The difference between PBA and CBA (the shaded area between the two
lines) has contracted since about 2008. This implies that the energy
content in imports is becoming more balanced over time. 

It is also apparent that PBA and CBA are closely correlated; an increase
or decline in one measure is followed by a similar change in the other.
Such a relationship between the two measures implies that it is just
a matter of time until a change experienced in one measure will be
replicated by the other. In other words; if PBA declines, CBA will
decline too and vice versa. For instance, CBA for BRIC in 2012 was
about the same as PBA for BRIC in 2010. For the EU28 and the US, these
changes are less visible because the rate of change in the two measures
is much slower. 

\section{Discussion and concluding remarks}

The aim of this paper has been to construct energy accounts for the
WIOD 2016 release and present the main trends in global energy footprints
for 2000 \textendash{} 2014, with a particular focus on the period
after 2009, for which the research on energy footprints is lacking. 

The newly constructed WIOD16 energy accounts were compared with the
existing WIOD13 energy accounts for the period 2000-2009, the period
for which the two databases overlap. This exercise shows the accuracy
of extended WIOD16 energy accounts. The results show that the difference
between WIOD16 and WIOD13 energy accounts for most countries (34 out
of 41) are within the 4\% range. The differences are mainly due to
the allocation procedure. Generally, such differences are not high
and in line with known differences between input-output results within
the IO community \citep{Moran2014}. 

To ensure that the two databases are comparable WIOD16 energy accounts
were calibrated to match those of WIOD13. As a result, PBA energy
is the same in both WIOD13 and WIOD16 energy accounts. However, as
shown in figure \ref{fig:CH4_Boxplot} CBA results differ across countries.
For most countries, the differences are within 5\% range, and in some
extreme cases, the differences range from -20\% to +40 \%. The negative
difference shows by how much CBA is underestimated and the positive
result shows how much it is overestimated. 

The exact source for these differences is not known, but few possible
explanations can be made. First, the differences can occur due to
different sectoral and spatial aggregation. WIOD13 is more aggregate
than WIOD 2016 both in terms of country and sector detail. The prevailing
view is that the finer the level of sector disaggregation the more
accurate the results. \citet{Su2010}use a single-country model to
investigate emissions embodied in exports of China and Singapore.
They suggest that around 40 sector aggregation is sufficient to capture
the majority of $CO_{2}$ emissions embodied in production. \citet{Bouwmeester2013}
show that for $CO_{2}$ emissions, aggregation errors are on average
2.3\% when sectoral detail is reduced from 129 to 59 sectors and about
3.4 \% when sectoral detail is reduced from 59 to 10 sectors. The
spatial aggregation error is on average 1.4\% when aggregating from
43 to 5 regions and 2.4\% when aggregating from 5 to 2 regions. However,
in most cases the results differ strongly across countries, suggesting
that a uniform prescription for the level of sectoral and spatial
detail is not possible. Interestingly, the countries that show the
largest aggregation error in \citet{Bouwmeester2013} study, also
appear as having the most significant differences between CBA estimates
(figure 5) of this study (e.g., the Baltic countries, Cyprus, Luxembourg,
Malta, Greece). 

Second, the differences can occur due to different accounting conventions.
The WIOD13 IO tables adhere to the 1993 version of System of National
Accounts (SNA), and the WIOD2016 release adhere to the 2008 version
of SNA. The SNA 2008 version involves two major changes in the recording
of international trade statistics. The first concerns changes of goods
sent abroad for processing and the second to merchanting \citep{VanDeVen2015}.
In the 1993 SNA goods sent abroad for processing and then returned
to the country from where they were dispatched are treated as undergoing
an effective change of ownership and recorded as imports and exports
\citet{Timmer2016}. The 2008 SNA version records transactions on
the basis of a change in (economic) ownership which means that goods
processed in one country on behalf of another are not recorded as
imports and exports even if they physically crossed the borders. These
changes have significant consequences for the input-output tables
and environmental analysis. Quantitively this leads to lower intermediate
consumption, output, import and export estimates. For some countries,
the reductions can be quite substantial \citet{Aspden2007}. \citet{VanRossum2014}show
that for the Netherlands changing from SNA 1993 to SNA 2008 lead to
-8.4\% lower estimates for emissions embodied in imports and +12.4\%
increase in the emission-trade balance. The authors conclude that
new SNA 2008 concepts undermine the potential of the environmental
input-output analysis.

Intensification of international trade and increasing production fragmentation
over the last few decades has made countries more interdependent on
one another's supply of resources. As shown in figure \ref{fig:CH4_PBA_CBA_global}
energy content embodied in trade remains high. PBA and CBA measures
are highly correlated. This has important implications for the decoupling
of energy use from economic growth debate. A prevailing hypothesis
suggests that the decoupling seen from the PBA perspective might be
a result of production outsourcing. One way to test this hypothesis
is to look at the CBA energy use which takes into account imports.
Figure \ref{fig:CH4_PBA_CBA_global} shows that the PBA and CBA measures
follow a similar trend, and a change in one is closely mirrored by
the other. This implies that the decoupling seen in the PBA case will
be reassembled by the CBA measure too. That is, PBA and CBA measures
will have a similar shape of the so-called Environmental Kuznets curve,
only the peak point will differ and the CBA will peak at a higher
point. 

\pagebreak{}

\bibliography{Chapter4}
\bibliographystyle{apa}

\pagebreak{}

\section*{\label{secCH4:Appendix-A}Appendix A\protect\pagebreak{}}

\begin{table}[H]
\begin{centering}
\thispagestyle{empty}
\par\end{centering}
\centering{}\setcounter{table}{0} \renewcommand{\thetable}{A.\arabic{table}}\caption{\label{tab:CH4_AppendixA_WIODCountry}World Input-Output Database
2016 Country Coverage }
\resizebox*{!}{\textheight}{%
\begin{tabular}{cll}
\toprule 
{\footnotesize{}No} & {\footnotesize{}Name} & {\footnotesize{}Code}\tabularnewline
\midrule
{\footnotesize{}1} & {\footnotesize{}Australia} & {\footnotesize{}AUS}\tabularnewline
{\footnotesize{}2} & {\footnotesize{}Austria} & {\footnotesize{}AUT}\tabularnewline
{\footnotesize{}3} & {\footnotesize{}Belgium} & {\footnotesize{}BEL}\tabularnewline
{\footnotesize{}4} & {\footnotesize{}Bulgaria} & {\footnotesize{}BGR}\tabularnewline
{\footnotesize{}5} & {\footnotesize{}Brazil} & {\footnotesize{}BRA}\tabularnewline
{\footnotesize{}6} & {\footnotesize{}Canada} & {\footnotesize{}CAN}\tabularnewline
{\footnotesize{}7} & {\footnotesize{}Switzerland} & {\footnotesize{}CHE}\tabularnewline
{\footnotesize{}8} & {\footnotesize{}People's Republic of China} & {\footnotesize{}CHN}\tabularnewline
{\footnotesize{}9} & {\footnotesize{}Cyprus} & {\footnotesize{}CYP}\tabularnewline
{\footnotesize{}10} & {\footnotesize{}Czech Republic} & {\footnotesize{}CZE}\tabularnewline
{\footnotesize{}11} & {\footnotesize{}Germany} & {\footnotesize{}DEU}\tabularnewline
{\footnotesize{}12} & {\footnotesize{}Denmark} & {\footnotesize{}DNK}\tabularnewline
{\footnotesize{}13} & {\footnotesize{}Spain} & {\footnotesize{}ESP}\tabularnewline
{\footnotesize{}14} & {\footnotesize{}Estonia} & {\footnotesize{}EST}\tabularnewline
{\footnotesize{}15} & {\footnotesize{}Finland} & {\footnotesize{}FIN}\tabularnewline
{\footnotesize{}16} & {\footnotesize{}France} & {\footnotesize{}FRA}\tabularnewline
{\footnotesize{}17} & {\footnotesize{}United Kingdom } & {\footnotesize{}GBR}\tabularnewline
{\footnotesize{}18} & {\footnotesize{}Greece} & {\footnotesize{}GRC}\tabularnewline
{\footnotesize{}19} & {\footnotesize{}Croatia} & {\footnotesize{}HRV}\tabularnewline
{\footnotesize{}20} & {\footnotesize{}Hungary} & {\footnotesize{}HUN}\tabularnewline
{\footnotesize{}21} & {\footnotesize{}Indonesia} & {\footnotesize{}IDN}\tabularnewline
{\footnotesize{}22} & {\footnotesize{}India} & {\footnotesize{}IND}\tabularnewline
{\footnotesize{}23} & {\footnotesize{}Ireland} & {\footnotesize{}IRL}\tabularnewline
{\footnotesize{}24} & {\footnotesize{}Italy} & {\footnotesize{}ITA}\tabularnewline
{\footnotesize{}25} & {\footnotesize{}Japan} & {\footnotesize{}JPN}\tabularnewline
{\footnotesize{}26} & {\footnotesize{}Republic of Korea} & {\footnotesize{}KOR}\tabularnewline
{\footnotesize{}27} & {\footnotesize{}Lithuania} & {\footnotesize{}LTU}\tabularnewline
{\footnotesize{}28} & {\footnotesize{}Luxembourg} & {\footnotesize{}LUX}\tabularnewline
{\footnotesize{}29} & {\footnotesize{}Latvia} & {\footnotesize{}LVA}\tabularnewline
{\footnotesize{}30} & {\footnotesize{}Mexico} & {\footnotesize{}MEX}\tabularnewline
{\footnotesize{}31} & {\footnotesize{}Malta} & {\footnotesize{}MLT}\tabularnewline
{\footnotesize{}32} & {\footnotesize{}Netherlands} & {\footnotesize{}NLD}\tabularnewline
{\footnotesize{}33} & {\footnotesize{}Norway} & {\footnotesize{}NOR}\tabularnewline
{\footnotesize{}34} & {\footnotesize{}Poland} & {\footnotesize{}POL}\tabularnewline
{\footnotesize{}35} & {\footnotesize{}Portugal} & {\footnotesize{}PRT}\tabularnewline
{\footnotesize{}36} & {\footnotesize{}Romania} & {\footnotesize{}ROU}\tabularnewline
{\footnotesize{}37} & {\footnotesize{}Russian Federation} & {\footnotesize{}RUS}\tabularnewline
{\footnotesize{}38} & {\footnotesize{}Slovakia} & {\footnotesize{}SVK}\tabularnewline
{\footnotesize{}39} & {\footnotesize{}Slovenia} & {\footnotesize{}SVN}\tabularnewline
{\footnotesize{}40} & {\footnotesize{}Sweden} & {\footnotesize{}SWE}\tabularnewline
{\footnotesize{}41} & {\footnotesize{}Turkey} & {\footnotesize{}TUR}\tabularnewline
{\footnotesize{}42} & {\footnotesize{}Taiwan} & {\footnotesize{}TWN}\tabularnewline
{\footnotesize{}43} & {\footnotesize{}United States} & {\footnotesize{}USA}\tabularnewline
{\footnotesize{}44} & {\footnotesize{}Rest of World} & {\footnotesize{}ROW}\tabularnewline
\bottomrule
\end{tabular}}
\end{table}

\begin{table}[H]
\begin{centering}
\thispagestyle{empty}
\par\end{centering}
\centering{}\setcounter{table}{1} \renewcommand{\thetable}{A.\arabic{table}}\caption{\label{tab:CH4Appendix_WIOD_Sector}World Input-Output Database 2016
Sectoral Coverage }
\resizebox*{!}{\textheight}{%
\begin{tabular}{c>{\raggedright}p{12cm}l}
\toprule 
{\footnotesize{}No} & {\footnotesize{}Name} & {\footnotesize{}Code}\tabularnewline
\midrule
{\footnotesize{}1} & {\footnotesize{}Crop and animal production, hunting and related service
activities} & {\footnotesize{}A01}\tabularnewline
{\footnotesize{}2} & {\footnotesize{}Forestry and logging} & {\footnotesize{}A02}\tabularnewline
{\footnotesize{}3} & {\footnotesize{}Fishing and aquaculture} & {\footnotesize{}A03}\tabularnewline
{\footnotesize{}4} & {\footnotesize{}Mining and quarrying} & {\footnotesize{}B}\tabularnewline
{\footnotesize{}5} & {\footnotesize{}Manufacture of food products, beverages and tobacco
products} & {\footnotesize{}C10-C12}\tabularnewline
{\footnotesize{}6} & {\footnotesize{}Manufacture of textiles, wearing apparel and leather
products} & {\footnotesize{}C13-C15}\tabularnewline
{\footnotesize{}7} & {\footnotesize{}Manufacture of wood and of products of wood and cork,
except furniture; manufacture of articles of straw and plaiting materials} & {\footnotesize{}C16}\tabularnewline
{\footnotesize{}8} & {\footnotesize{}Manufacture of paper and paper products} & {\footnotesize{}C17}\tabularnewline
{\footnotesize{}9} & {\footnotesize{}Printing and reproduction of recorded media} & {\footnotesize{}C18}\tabularnewline
{\footnotesize{}10} & {\footnotesize{}Manufacture of coke and refined petroleum products } & {\footnotesize{}C19}\tabularnewline
{\footnotesize{}11} & {\footnotesize{}Manufacture of chemicals and chemical products } & {\footnotesize{}C20}\tabularnewline
{\footnotesize{}12} & {\footnotesize{}Manufacture of basic pharmaceutical products and pharmaceutical
preparations} & {\footnotesize{}C21}\tabularnewline
{\footnotesize{}13} & {\footnotesize{}Manufacture of rubber and plastic products} & {\footnotesize{}C22}\tabularnewline
{\footnotesize{}14} & {\footnotesize{}Manufacture of other non-metallic mineral products} & {\footnotesize{}C23}\tabularnewline
{\footnotesize{}15} & {\footnotesize{}Manufacture of basic metals} & {\footnotesize{}C24}\tabularnewline
{\footnotesize{}16} & {\footnotesize{}Manufacture of fabricated metal products, except machinery
and equipment} & {\footnotesize{}C25}\tabularnewline
{\footnotesize{}17} & {\footnotesize{}Manufacture of computer, electronic and optical products} & {\footnotesize{}C26}\tabularnewline
{\footnotesize{}18} & {\footnotesize{}Manufacture of electrical equipment} & {\footnotesize{}C27}\tabularnewline
{\footnotesize{}19} & {\footnotesize{}Manufacture of machinery and equipment n.e.c.} & {\footnotesize{}C28}\tabularnewline
{\footnotesize{}20} & {\footnotesize{}Manufacture of motor vehicles, trailers and semi-trailers} & {\footnotesize{}C29}\tabularnewline
{\footnotesize{}21} & {\footnotesize{}Manufacture of other transport equipment} & {\footnotesize{}C30}\tabularnewline
{\footnotesize{}22} & {\footnotesize{}Manufacture of furniture; other manufacturing} & {\footnotesize{}C31\_C32}\tabularnewline
{\footnotesize{}23} & {\footnotesize{}Repair and installation of machinery and equipment} & {\footnotesize{}C33}\tabularnewline
{\footnotesize{}24} & {\footnotesize{}Electricity, gas, steam and air conditioning supply} & {\footnotesize{}D35}\tabularnewline
{\footnotesize{}25} & {\footnotesize{}Water collection, treatment and supply} & {\footnotesize{}E36}\tabularnewline
{\footnotesize{}26} & {\footnotesize{}Sewerage; waste collection, treatment and disposal
activities; materials recovery; remediation activities and other waste
management services } & {\footnotesize{}E37-E39}\tabularnewline
{\footnotesize{}27} & {\footnotesize{}Construction} & {\footnotesize{}F}\tabularnewline
{\footnotesize{}28} & {\footnotesize{}Wholesale and retail trade and repair of motor vehicles
and motorcycles} & {\footnotesize{}G45}\tabularnewline
{\footnotesize{}29} & {\footnotesize{}Wholesale trade, except of motor vehicles and motorcycles} & {\footnotesize{}G46}\tabularnewline
{\footnotesize{}30} & {\footnotesize{}Retail trade, except of motor vehicles and motorcycles} & {\footnotesize{}G47}\tabularnewline
{\footnotesize{}31} & {\footnotesize{}Land transport and transport via pipelines} & {\footnotesize{}H49}\tabularnewline
{\footnotesize{}32} & {\footnotesize{}Water transport} & {\footnotesize{}H50}\tabularnewline
{\footnotesize{}33} & {\footnotesize{}Air transport} & {\footnotesize{}H51}\tabularnewline
{\footnotesize{}34} & {\footnotesize{}Warehousing and support activities for transportation} & {\footnotesize{}H52}\tabularnewline
{\footnotesize{}35} & {\footnotesize{}Postal and courier activities} & {\footnotesize{}H53}\tabularnewline
{\footnotesize{}36} & {\footnotesize{}Accommodation and food service activities} & {\footnotesize{}I}\tabularnewline
{\footnotesize{}37} & {\footnotesize{}Publishing activities} & {\footnotesize{}J58}\tabularnewline
{\footnotesize{}38} & {\footnotesize{}Motion picture, video and television programme production,
sound recording and music publishing activities; programming and broadcasting
activities} & {\footnotesize{}J59\_J60}\tabularnewline
{\footnotesize{}39} & {\footnotesize{}Telecommunications} & {\footnotesize{}J61}\tabularnewline
{\footnotesize{}40} & {\footnotesize{}Computer programming, consultancy and related activities;
information service activities} & {\footnotesize{}J62\_J63}\tabularnewline
{\footnotesize{}41} & {\footnotesize{}Financial service activities, except insurance and
pension funding} & {\footnotesize{}K64}\tabularnewline
{\footnotesize{}42} & {\footnotesize{}Insurance, reinsurance and pension funding, except
compulsory social security} & {\footnotesize{}K65}\tabularnewline
{\footnotesize{}43} & {\footnotesize{}Activities auxiliary to financial services and insurance
activities} & {\footnotesize{}K66}\tabularnewline
{\footnotesize{}44} & {\footnotesize{}Real estate activities} & {\footnotesize{}L68}\tabularnewline
{\footnotesize{}45} & {\footnotesize{}Legal and accounting activities; activities of head
offices; management consultancy activities} & {\footnotesize{}M69\_M70}\tabularnewline
{\footnotesize{}46} & {\footnotesize{}Architectural and engineering activities; technical
testing and analysis} & {\footnotesize{}M71}\tabularnewline
{\footnotesize{}47} & {\footnotesize{}Scientific research and development} & {\footnotesize{}M72}\tabularnewline
{\footnotesize{}48} & {\footnotesize{}Advertising and market research} & {\footnotesize{}M73}\tabularnewline
{\footnotesize{}49} & {\footnotesize{}Other professional, scientific and technical activities;
veterinary activities} & {\footnotesize{}M74\_M75}\tabularnewline
{\footnotesize{}50} & {\footnotesize{}Administrative and support service activities} & {\footnotesize{}N}\tabularnewline
{\footnotesize{}51} & {\footnotesize{}Public administration and defence; compulsory social
security} & {\footnotesize{}O84}\tabularnewline
{\footnotesize{}52} & {\footnotesize{}Education} & {\footnotesize{}P85}\tabularnewline
{\footnotesize{}53} & {\footnotesize{}Human health and social work activities} & {\footnotesize{}Q}\tabularnewline
{\footnotesize{}54} & {\footnotesize{}Other service activities} & {\footnotesize{}R\_S}\tabularnewline
{\footnotesize{}55} & {\footnotesize{}Activities of households as employers; undifferentiated
goods- and services-producing activities of households for own use} & {\footnotesize{}T}\tabularnewline
{\footnotesize{}56} & {\footnotesize{}Activities of extraterritorial organizations and bodies} & {\footnotesize{}U}\tabularnewline
{\footnotesize{}57} & {\footnotesize{}Households} & {\footnotesize{}HH}\tabularnewline
\bottomrule
\end{tabular}}
\end{table}

\thispagestyle{empty}

\begin{table}[H]
\setcounter{table}{2} \renewcommand{\thetable}{A.\arabic{table}}
\centering{}\caption{IEA and WIOD2016 Energy product correspondence }
\resizebox*{!}{\textwidth}{%
\begin{tabular}{cl>{\raggedright}p{8cm}}
\toprule 
{\footnotesize{}No} & {\footnotesize{}WIOD16 energy } & {\footnotesize{}IEA energy Product}\tabularnewline
\midrule
{\footnotesize{}1} & {\footnotesize{}HCOAL} & {\footnotesize{}Anthracite; Coking coal; Other bituminous coal; Sub-bituminous
coal; Patent fuel}\tabularnewline
{\footnotesize{}2} & {\footnotesize{}BCOAL} & {\footnotesize{}Lignite; Coal tar; BKB; Peat; Peat products; Oil shale
and oil sands}\tabularnewline
{\footnotesize{}3} & {\footnotesize{}COKE} & {\footnotesize{}Coke oven coke; Gas coke}\tabularnewline
{\footnotesize{}4} & {\footnotesize{}CRUDE} & {\footnotesize{}Crude oil; Natural gas liquids; Refinery feedstocks;
Additives/blending components; Other hydrocarbons}\tabularnewline
{\footnotesize{}5} & {\footnotesize{}DIESEL} & {\footnotesize{}Gas/diesel oil excl. biofuels}\tabularnewline
{\footnotesize{}6} & {\footnotesize{}GASOLINE} & {\footnotesize{}Motor gasoline excl. biofuels}\tabularnewline
{\footnotesize{}7} & {\footnotesize{}JETFUEL} & {\footnotesize{}Aviation gasoline; Gasoline type jet fuel; Kerosene
type jet fuel excl. biofuels}\tabularnewline
{\footnotesize{}8} & {\footnotesize{}LFO} & \tabularnewline
{\footnotesize{}9} & {\footnotesize{}HFO} & {\footnotesize{}Fuel oil}\tabularnewline
{\footnotesize{}10} & {\footnotesize{}NAPHTA} & {\footnotesize{}Naphtha}\tabularnewline
{\footnotesize{}11} & {\footnotesize{}OTHPETRO} & {\footnotesize{}Refinery gas; Ethane; Liquefied petroleum gases (LPG);
Other kerosene; White spirit \& SBP; Lubricants; Bitumen; Paraffin
waxes; Petroleum coke; Other oil products}\tabularnewline
{\footnotesize{}12} & {\footnotesize{}NATGAS} & {\footnotesize{}Natural gas}\tabularnewline
{\footnotesize{}13} & {\footnotesize{}OTHGAS} & {\footnotesize{}Gas works gas; Coke oven gas; Blast furnace gas; Other
recovered gases}\tabularnewline
{\footnotesize{}14} & {\footnotesize{}WASTE} & {\footnotesize{}Industrial waste; Municipal waste (renewable); Municipal
waste (non-renewable)}\tabularnewline
{\footnotesize{}15} & {\footnotesize{}BIOGASOL} & {\footnotesize{}Biogasoline; Other liquid biofuels}\tabularnewline
{\footnotesize{}16} & {\footnotesize{}BIODIESEL} & {\footnotesize{}Biodiesels}\tabularnewline
{\footnotesize{}17} & {\footnotesize{}BIOGAS} & {\footnotesize{}Biogases}\tabularnewline
{\footnotesize{}18} & {\footnotesize{}OTHRENEW} & {\footnotesize{}Primary solid biofuels; Charcoal}\tabularnewline
{\footnotesize{}19} & {\footnotesize{}ELECTR} & {\footnotesize{}Electricity}\tabularnewline
{\footnotesize{}20} & {\footnotesize{}HEATPROD} & {\footnotesize{}Elec/heat output from non-specified manufactured gases;
Heat output from non-specified combustible fuels; Heat}\tabularnewline
{\footnotesize{}21} & {\footnotesize{}NUCLEAR} & {\footnotesize{}Nuclear}\tabularnewline
{\footnotesize{}22} & {\footnotesize{}HYDRO} & {\footnotesize{}Hydro }\tabularnewline
{\footnotesize{}23} & {\footnotesize{}GEOTHERM} & {\footnotesize{}Geothermal }\tabularnewline
{\footnotesize{}24} & {\footnotesize{}SOLAR} & {\footnotesize{}Solar photovoltaics; Solar thermal}\tabularnewline
{\footnotesize{}25} & {\footnotesize{}WIND} & {\footnotesize{}Wind}\tabularnewline
{\footnotesize{}26} & {\footnotesize{}OTHSOURC} & {\footnotesize{}Tide, wave and ocean; Other sources}\tabularnewline
{\footnotesize{}27} & {\footnotesize{}LOSS} & \tabularnewline
\bottomrule
\end{tabular}}
\end{table}

\pagebreak{}
\end{document}